\documentclass{article}
\usepackage{spconf,amsmath,graphicx}
\usepackage{multirow}
\usepackage{xcolor}
\usepackage[colorlinks,linkcolor=blue]{hyperref}

\title{Mix-Domain Contrastive Learning for Unpaired H\&E-to-IHC Stain Translation}
\name{Song Wang\textsuperscript{1,2}, Zhong Zhang\textsuperscript{2} , Huan Yan\textsuperscript{3}, Ming Xu\textsuperscript{4}, Guanghui Wang\textsuperscript{1}\thanks{Corresponding Author: Guanghui Wang(wangcs@torontomu.ca).\\ This work was partly supported by the Natural Sciences and Engineering Research Council of Canada (NSERC).}}
\address{\textsuperscript{1} Toronto Metropolitan University, Toronto, M5B 2K3, Canada\\
\textsuperscript{2} Zhengzhou University, Zhengzhou, 450001, China\\
\textsuperscript{3} The First Affiliated Hospital of Zhengzhou University, Zhengzhou, 450000, China\\ \textsuperscript{4} Zhongyuan University of Technology, Zhengzhou, 451191, China}
\begin{document}
%
\maketitle
\begin{abstract}
H\&E-to-IHC stain translation techniques offer a promising solution for precise cancer diagnosis, especially in low-resource regions where there is a shortage of health professionals and limited access to expensive equipment. Considering the pixel-level misalignment of H\&E-IHC image pairs, current research explores the pathological consistency between patches from the same positions of the image pair. However, most of them overemphasize the correspondence between domains or patches, overlooking the side information provided by the non-corresponding objects. In this paper, we propose a Mix-Domain Contrastive Learning (MDCL) method to leverage the supervision information in unpaired H\&E-to-IHC stain translation. Specifically, the proposed MDCL method aggregates the inter-domain and intra-domain pathology information by estimating the correlation between the anchor patch and all the patches from the matching images, encouraging the network to learn additional contrastive knowledge from mixed domains. With the mix-domain pathology information aggregation, MDCL enhances the pathological consistency between the corresponding patches and the component discrepancy of the patches from the different positions of the generated IHC image. Extensive experiments on two H\&E-to-IHC stain translation datasets, namely MIST and BCI, demonstrate that the proposed method achieves state-of-the-art performance across multiple metrics. Code is available at \url{https://github.com/SSongWang/Mix-DomainContrastiveLearning.git}.
\end{abstract}
\section{Introduction}
Hematoxylin and Eosin (H\&E) staining is the most representative histochemical staining technique in the histopathological analysis workflow, enabling the visualization of distinct tissue components in various colors, thereby facilitating cancer identification \cite{bentaieb2017adversarial,lin2022unpaired,zeng2022semi}. Due to its cost-effectiveness, H\&E staining is widely utilized in cancer diagnosis \cite{chen2017computer}. However, relying solely on H\&E staining images may not suffice for accurate differentiation of cancer subtypes by pathologists. To improve this limitation, Immunohistochemistry (IHC) staining techniques are developed to visualize specific antigens or proteins in the tissue via the interaction between labeled antibodies and intracellular antigens \cite{ramos2005technical}. For example, Ki-67 IHC examination can stain positive tumor cells brown and negative ones blue, providing a selective and high-contrast visual result for pathologists to confirm the malignancy types \cite{sheikh2003correlation}. Despite its high accuracy, IHC examination is labor-intensive and time-consuming, often requiring expensive equipment and advanced technologies. This poses a significant barrier to the widespread applications of pathology diagnostic services in low-income regions. Therefore, developing an effective and affordable method for IHC image acquisition is imperative.

\begin{figure}[tbp]
\includegraphics[width=0.49\textwidth]{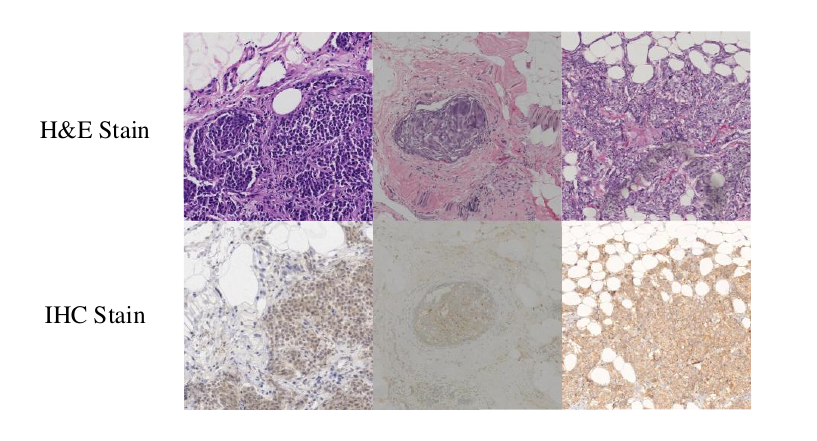}
\caption{Some examples of H\&E-IHC image pairs from public MIST \cite{li2023adaptive} and BCI \cite{liu2022bci} datasets. Matching H\&E-IHC images are unpaired at the pixel level.} \label{he-ihc}
\end{figure}

To achieve this objective, researchers exploit data-driven approaches to transfer H\&E-stained images to IHC-stained ones. However, an outstanding challenge faced by the approaches is how to build matching H\&E-IHC image pairs. Since a slice can be only stained once, obtaining pixel-aligned H\&E-IHC image pairs is physically infeasible. In this case, the second best choice is cutting the serial tissue sections from the same tissue block and staining them separately. The adjacent cuts stained by different methods are viewed as the matching H\&E-IHC image pairs. Several matching H\&E-IHC image pairs from available public datasets are shown in Fig. \ref{he-ihc}. It is obvious that these matching H\&E-IHC image pairs are pixel-wise unpaired. For the unpaired matching images, there are some attempts to register them \cite{zeng2022semi,liu2020predict,liu2022bci}. For example, Liu et al. \cite{liu2022bci} propose to register the matching H\&E-IHC images at the structure level and obtain the structure-aligned H\&E-IHC image pairs. However, those registration methods suffer from expert shortages and low efficiency. 


An alternative solution is training the H\&E-to-IHC translation model by directly utilizing the unpaired images. The most representative H\&E-to-IHC translation methods are the CycleGan-based ones \cite{zeng2022semi,xu2021domain,zhang2021six,liu2021unpaired,xu2021drb}, where the inverse generation process ensures the structure consistency between the generated image and the source image. However, these approaches hardly generate virtual IHC images with the same pathological properties as input H\&E images. To address this issue, patch-wise contrastive learning is introduced into H\&E-to-IHC stain translation. Li et al. \cite{li2023adaptive} build a contrastive learning translation framework to maintain the pathological consistency between the corresponding patches from different images by mapping them together in embedding spaces. 

\begin{figure}[htb]
\includegraphics[width=0.46\textwidth]{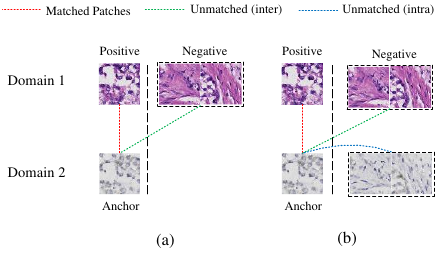}
\caption{The difference between (a) existing methods and (b) MDCL. For each anchor patch, existing methods only consider inter-domain patches (red and green). MDCL utilizes patches from both inter- (red and green) and intra-domain (blue).} \label{difference}
\end{figure}

\textit{The implicit hypothesis behind patch-wise contrastive learning is that the corresponding patches from matching images should have the same probability of cancer diagnosis.} Based on this hypothesis, patch-wise contrastive learning has shown promising achievements in unpaired image-to-image translation tasks. However, the patch-wise contrastive learning method overemphasizes the contribution of the patches from the inter-domain (Fig. \ref{difference}(a)), ignoring the patches from the intra-domain (Fig. \ref{difference}(b)). Since the cancer cells distribute randomly in the tissue, as the brown cells in the IHC images from Fig. \ref{he-ihc}, we argue that the patches from the intra-domain typically have a different probability of cancer diagnosis but a similar appearance with the anchor, which provides excellent hard negatives for the contrastive learning. By collecting the patches from the intra-domain as negatives, the generation network is enforced to focus on the subtle differences between the patches from the same domain, further generating the virtual IHC-stained images with discriminative component details.

In this paper, we propose a Mix-Domain Contrastive Learning (MDCL) method for unpaired H\&E-to-IHC stain translation. Specifically, MDCL method estimates the correlation between the anchor patch and all the patches from both inter- and intra-domains. This distinctive feature encourages the network to acquire additional contrastive knowledge from both staining domains and generate the virtual IHC-stained images with discriminative local details. 

The main contributions of this paper include:\\[-18pt]
\begin{itemize}\setlength\itemsep{1pt}
  \item MDCL is proposed for the unpaired translation of H\&E to IHC stains. MDCL combines inter-domain and intra-domain pathology information to improve pathological consistency between the input-generated image pairs and the component discrepancies in patches from different positions of the generated IHC image.
 
  \item Extensive experimental evaluations on two H\&E-to-IHC stain translation datasets, MIST and BCI, demonstrate that the proposed method achieves state-of-the-art performance. Visualization results further illustrate MDCL's superiority in preserving component details. 
\end{itemize}

\section{Methods}
In this section, we first present the H\&E-to-IHC stain translation framework based on patch-wise constrastive learning. Then, we provide a detailed presentation of the mix-domain contrastive learning method in Sec. 2.2.

\begin{figure*}[!t]
\centering
\includegraphics[width=0.95\textwidth]{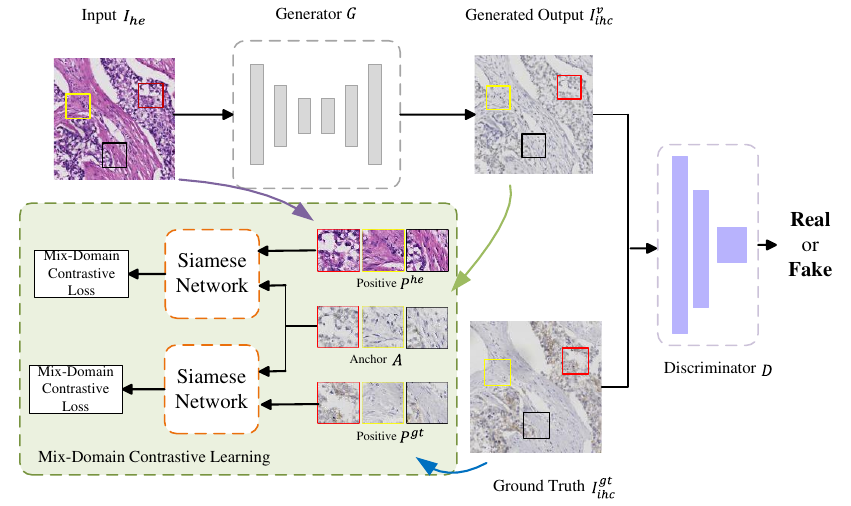}
\caption{H\&E-to-IHC stain translation learning framework. Given the input H\&E stained image $I_{he}$, the generator $G$ generates the virtual IHC stained image $I_{ihc}^v$. $M$ patches from $I_{ihc}^v$ are randomly sampled to construct the anchor set $A$. At the same locations, the positive patches $p^{he}_i$ from $I_{he}$ and $p^{gt}_i$ from $I_{ihc}^{gt}$ are selected to build $P^{he}$ and $P^{gt}$. The Siamese network $f$ maps each pair of $(a_i, p_i)$ together under the constraint of the mix-domain contrastive loss.} \label{framework}
\end{figure*}

\subsection{H\&E-to-IHC Stain Translation Learning} \label{framwork}
Fig. \ref{framework} illustrates the proposed H\&E-to-IHC stain translation learning framework. Given the input H\&E stained image $I_{he}$, its corresponding virtual IHC stained image $I_{ihc}^v$ is generated by the generator $G$ of GAN. The discriminator $D$ is utilized to distinguish whether the virtual IHC stained image $I_{ihc}^v$ is similar to the real one $I_{ihc}^{gt}$. For the generated $I_{ihc}^v$, $M$ patches are randomly sampled to construct the anchor set $A=\left\{a_1,a_2, \cdots, a_M\right\}$. Meanwhile, the patches at the same locations are selected from the input H\&E image $I_{he}$ as the positive set $P^{he}=\left\{p_1^{he},p_2^{he}, \cdots, p_M^{he}\right\}$. For the matching pairs $(a_i, p_i^{he})_{i=1,\cdots, M}$, a Siamese network $f$ is built to map the matching patches together under the constraint of the constrastive loss. To exploit the supervised information from the ground-truth $I_{ihc}^{gt}$, a pixel-misaligned positive set $P^{gt}=\left\{p_1^{gt},p_2^{gt}, \cdots, p_M^{gt}\right\}$ is constructed, where $p_i^{gt}$ is the corresponding patch to the anchor $a_i$. Similar to $p_i^{he}$, $p_i^{gt}$ is also demanded to be closer to $a_i$ in the embedding space. To relieve the misalignment problem at the pixel level, the adaptive weighting scheme from \cite{li2023adaptive} is adopted to control the influence of patch pairs on the loss value.

\subsection{Mix-Domain Contrastive Learning}
Contrastive learning aims to pull the positive patch pairs closer and push the negative patch pairs apart in the embedding space. We take the input-generated image pair as an example in the following explanation. For each virtual IHC patch $a_i$, its embedding is denoted as $z_i = f(a_i)$. Similarly, the embedding of the corresponding H\&E patch $p_i^{he}$ can be represented as $z_i^{he} = f(p_i^{he})$. It is important to note that the last layer of the Siamese network $f$ is typically a $L_2$ Normalization layer, allowing the patches to be mapped onto a unit hypersphere.

For the anchor set $A$ and its positive set $P^{he}$, the probability of $a_i$ matching with $p_i^{he}$ can be represented as
\begin{equation}\label{eq_prob}
    \mathcal{P}_i^{he}=\frac{\exp(z_i \cdot z_i^{he}/\tau)}{\exp(z_i \cdot z_i^{he} /\tau) + \sum_{j=1,j\neq i}^{M} \exp(z_i \cdot z_j^{he}/\tau)},
\end{equation}
where $\tau$ stands for the temperature which controls the concentration level of the class distribution \cite{wu2018unsupervised}. In Eq.(\ref{eq_prob}), $z_i^{he}$ denotes the positive embedding of the anchor $z_i$, while $z^{he}_{j_{j\neq i}}$ represents the negative embedding. Since $z$ is on the unit hypersphere, the dot product of $z_i$ and $z_j^{he}$ stands for the cosine similarity between the anchor patch from IHC domain and the $j$-th patch from H\&E domain. The ratio of $z_i \cdot z_i^{he}$ to $\sum_{j=1}^M z_i \cdot z_j^{he}$ indicates the probability of $a_i$ matching with $p_i^{he}$. The original patch-based contrastive loss \cite{park2020contrastive} is designed as the negative log-likelihood of the probability $\mathcal{P}_i$
\begin{equation}\label{eq_nce}
\begin{aligned}
    \mathcal{L}_i & = -\log\left[ \frac{\exp(z_i \cdot z_i^{he}/\tau)}{\exp(z_i \cdot z_i^{he} /\tau) + \sum_{j=1,j\neq i}^{M} \exp(z_i \cdot z_j^{he}/\tau)} \right].
\end{aligned}
\end{equation}

Minimizing Eq.(\ref{eq_nce}) can be viewed as maximizing the probability of $a_i$ matching with $p_i^{he}$.

It is observed that the existing patch-based contrastive loss in Eq.(\ref{eq_nce}) only considers the relationship between the anchor and the patches from different domains, like Fig. \ref{difference}(a). Following the implicit hypothesis behind patch-wise contrastive learning, it is well-known that the non-corresponding patches have different probabilities of cancer diagnosis. This hypothesis inspires us that the patches from the same domain with the anchor can also be viewed as negative patches, like Fig.\ref{difference}(b). In this case, the mix-domain contrastive loss for anchor $a_i$ can be formulated as

\begin{small}
\footnotesize
   \begin{equation}\label{eq_icl}
   \begin{aligned}
    \mathcal{L}_i^{he} = -\log  \left[ \frac{\exp(z_i \cdot z_i^{he}/\tau)}{\sum_{j=1}^{M} \exp(z_i \cdot z_j^{he}/\tau)+\sum_{j=1,j\neq i}^{M} \exp(z_i \cdot z_j/\tau)}  \right],
   \end{aligned}
   \end{equation} 
\end{small}


\noindent where the first item in the denominator represents the inter-domain similarity while the intra-domain correlation is explored by the second item. The symbol $z_j$ denotes the embedding of the negative patch from the same domain. By introducing negative patches with similar colors and appearances, the mix-domain contrastive loss aims at discriminating the subtle difference between the patches from the same domain, which in turn encourages the generator $G$ to pay more attention to detail generation. The mix-domain contrastive loss for the anchor set $A$ and the positive set $P^{he}$ is written as follows
\begin{equation}\label{eq_L}
    L^{he}_{mix}=\sum_{i=1}^M \mathcal{L}_{i}^{he}.
\end{equation}

\begin{figure}[htb]
\centering
\includegraphics[width=0.45\textwidth]{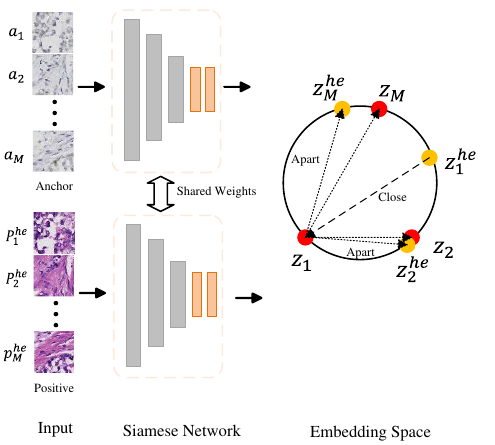}
\caption{Mix-domain contrastive learning for the anchor set $A$ and positive set $P^{he}$. All the patches from $A$ and $P^{he}$ are at first mapped onto the unit hypersphere by the Siamese network. For each anchor $a_i$, its embedding $z_i$ is demanded to be close to $z_i^{he}$ (the embedding of the positive $p_i^{he}$) and far away from all the rest.} \label{Contrastive}
\end{figure}

The whole learning process is illustrated in Fig.\ref{Contrastive}. For the anchor set $A$ and positive set $P^{he}$. All the patches from $A$ and $P^{he}$ are mapped onto the unit hypersphere by the Siamese network. For each anchor patch $a_i$, the mix-domain contrastive loss demands it to be close to the positive patch $p_i^{he}$ and far away from all the rest of the patches on the unit hypersphere.

Similarly,  the mix-domain contrastive loss for the anchor set $A$ and the positive set $P^{gt}$ is formulated as
\begin{equation}\label{eq_Lgt}
    L^{gt}_{mix}=\sum_{i=1}^M \omega_i \mathcal{L}_{i}^{gt},
\end{equation}
where $\omega_i$ means the adaptive weight \cite{li2023adaptive} for the constrastive loss generated by the anchor $a_i$. In Eq.\ref{eq_Lgt}, the groundtruth image and the generated IHC image are viewed as different domains. 

To sum up, the overall learning objective for the generator $G$ is defined as
\begin{equation}\label{eq_overal}
 L_{overall}=L_{adv}+L^{he}_{mix}+L^{gt}_{mix}+\lambda_{GP}L_{GP},
\end{equation}
where $L_{adv}$ stands for the standard adversarial loss generated by the generator whereas $L_{GP}$ represents the Gaussian Pyramid-based reconstruction loss \cite{liu2022bci}. The coefficient $\lambda_{GP}$ is utilized to balance the contributions of $L_{GP}$ to the generator training.

\section{Experiments}
\subsection{Datasets}
\textbf{MIST.} The Multi-IHC Stain Translation dataset \cite{li2023adaptive} consists of four subsets with different IHC stain approaches, namely HER2, ER, Ki67, and PR. For each subset, there are more than 4,000 image pairs for training and 1,000 image pairs for testing. The resolution of all images is 1024$\times$1024.

\noindent\textbf{BCI.} The Breast Cancer Immunohistochemical dataset \cite{liu2022bci} includes 4,873 H\&E-IHC image pairs where 3,896 pairs are selected as training set and the rest 977 pairs as testing set by following the default setting\footnote{https://bupt-ai-cz.github.io/BCI/}. The IHC-stained images have four HER2 expression statuses: IHC 0, IHC 1+, IHC 2+, and IHC 3+, standing for the different staining stages. All images in BCI are of size 1024$\times$1024.

\begin{table*}[!t]
\caption{Evaluation results on MIST. The best values are in bold. KID values are shown by multiplying 1000.}\label{tab_mist}
\centering
\begin{tabular}{|c|l|c|ccccc|c|}
\hline
\multirow{2}*{Subset} &  \multirow{2}*{Method} & \multirow{2}*{FID$\downarrow$} & \multicolumn{5}{c|}{PHV$_{T=0.01}$$\downarrow$} & \multirow{2}*{KID$\downarrow$}\\
& & & layer1 & layer2 & layer3 & layer4 & average& \\
\hline
\multirow{6}*{HER2} &  CycleGAN & 240.3& 0.5633 & 0.6346 & 0.4695 & 0.8871 & 0.6386 & 311.1\\
& CUT+$L_{GP}$ & 66.8 & 0.5321 & 0.4826 & 0.3060 & 0.8323 & 0.5383&19.0\\
& Pix2Pix& 137.3 & 0.5516 & 0.5070 & 0.3253 & 0.8511 & 0.5588& 82.9 \\
& PyramidP2P & 104.0 & 0.4787 & 0.4524 & 0.3313 & 
 0.8423 & 0.5262 & 61.8 \\
& ASP & 51.4 & 0.4534 & 0.4150 & 0.2665 & 0.8174 & 0.4881 & 12.4\\
& MDCL & {\bfseries 44.4} & {\bfseries 0.4371} & {\bfseries 0.3944} & {\bfseries 0.2518} & {\bfseries 0.8171}& {\bfseries 0.4751} &{\bfseries7.5}\\
\hline
\multirow{6}*{ER2} &  CycleGAN & 125.7& 0.5175 & 0.5092 &0.3710 &0.8672 &0.5662 & 95.1\\
& CUT+$L_{GP}$ & 43.7 & 0.4531 & 0.4079 & 0.2725 &0.8194 &0.4882& 8.7\\
& Pix2Pix& 128.1&  0.5818 & 0.5282 & 0.3700 &0.8620 &0.5855& 79.0 \\
& PyramidP2P & 107.4 &  0.4767& 0.4538& 0.3757 &0.8567& 0.5407 & 84.2 \\
& ASP & 41.4 & {\bfseries 0.4336} &0.4007 &0.2649 &{\bfseries0.8205} &{\bfseries 0.4799} & 5.8\\
& MDCL & {\bfseries 34.9} &  0.4533 & {\bfseries  0.3969} & {\bfseries 0.2646} &  0.8238&  0.4854 &{\bfseries 3.6}\\
\hline
\multirow{6}*{Ki67} &  CycleGAN & 343.9& 0.8274 &0.8275 &0.6081 &0.9038 &0.7917 & 317.9\\
& CUT+$L_{GP}$ & 76.1 & 0.5426 &0.4739 &0.3160 &0.8415 &0.5435 &43.5\\
& Pix2Pix& 147.0 & 0.5468 &0.4905 &0.3415 &0.8496 &0.5571& 142.4 \\
& PyramidP2P & 94.4 & 0.4533 &0.4222 &0.3360 &0.8363 &0.5120 & 78.0 \\
& ASP & 51.0 & 0.4472 &0.4001 &0.2701 &0.8128 &0.4826& 19.1\\
& MDCL & {\bfseries 30.8} & {\bfseries 0.4005} & {\bfseries 0.3638} & {\bfseries 0.2465} & {\bfseries 0.8064}& {\bfseries 0.4543} &{\bfseries6.1}\\
\hline
\multirow{6}*{PR} &  CycleGAN & 96.1& 0.5334 &0.5554 &0.3867 &0.8654 &0.5852& 96.6\\
& CUT+$L_{GP}$ & 54.6 & 0.4656 &0.4128 &0.2724 &0.8154 &0.4916&20.1\\
& Pix2Pix& 183.8 & 0.6027 &0.5569 &0.4043& 0.8601 &0.6060& 148.1 \\
& PyramidP2P & 98.8 & 0.5078 &0.4682& 0.3509 &0.8446 &0.5429& 59.5 \\
& ASP & 44.8 & 0.4484 &0.3898& 0.2564 &0.8080 &0.4757 & 10.2\\
& MDCL & {\bfseries 38.3} & {\bfseries 0.4320} & {\bfseries 0.3768} & {\bfseries 0.2499} & {\bfseries 0.8052}& {\bfseries 0.4660} &{\bfseries7.1}\\
\hline
\end{tabular}
\end{table*}

\subsection{Implementation Details}
Following the setting of \cite{li2023adaptive}, ResNet-6Blocks \cite{johnson2016perceptual} and PatchGAN \cite{isola2017image} are employed to be the generator and discriminator, respectively. The Siamese network is composed of the encoder part of the generator, two MLP layers with a ReLU layer in the middle, and an L2 Normalization layer. During the training, the images are randomly cropped to be 512$\times$512. The training epoch is set to 40. The initial learning rate is $2\times 10^{-4}$ and decreases using linear decay after 30 epochs. The Adam optimizer \cite{kingma2014adam} with $\beta_1=0.5$ and $\beta_2=0.999$ is utilized for training. For the overall loss in Eq.(\ref{eq_overal}), the coefficient $\lambda_{GP}$ is set to 10, same as in \cite{li2023adaptive}. The ``lambda\_linear'' scheme is chosen in Eq.(\ref{eq_Lgt}) to weight the pix-misaligned patches. For each image pair, 256 patches are randomly selected to generate the anchor and positive sets.

Multiple metrics are employed to measure the performance of the proposed method. At first, this paper employs Fréchet Inception Distance (FID) \cite{heusel2017gans} to compare the distributions of generated and groundtruth IHC images in the InceptionV3 feature space. By relaxing the Gaussian assumption in FID, Kernel Inception Distance (KID) \cite{binkowski2018demystifying} measures the squared Maximum Mean Discrepancy (MMD) between the Inception representations of the generated and groundtruth images. Additionally, Perceptual Hash Value (PHV) \cite{liu2021unpaired} is employed to gauge the content similarity of the generated and groundtruth pairs. In all the aforementioned metrics, a lower value means a better performance.

\subsection{Experiment on MIST}
\textbf{Evaluation Results}. We compared the proposed method with the existing image-to-image translation methods which include CycleGAN \cite{zhu2017unpaired}, CUT \cite{park2020contrastive}, Pix2Pix \cite{isola2017image}, PyramidP2P \cite{liu2022bci}, ASP \cite{li2023adaptive}. The quantitative results are shown in Table \ref{tab_mist}. We can see from the results that the lowest FID and KID are achieved by MDCL on all the subsets, demonstrating the superiority of MDCL on different H\&E-IHC translation tasks. Moreover, MDCL obtains the best PHV on most subsets, except for ER2 subset where the performance of MDCL is slightly worse than ASP.

\noindent\textbf{Ablation Study.} To verify the effectiveness of MDCL, this paper performs an ablation study and compares it with the original patch-based contrastive loss. The results on HER2 subset are presented in Table \ref{tab_abl}. We first evaluate the influence of MDCL only on the generated-input patch pair loss $L^{he}$. From the first two rows of Table \ref{tab_abl}, we can see that MDCL improves the H\&E-to-IHC translation performance across all three metrics. We also assess the effect of MDCL on the combination of $L^{he}$ and $L^{gt}$, where $L^{gt}$ denotes the patch-based contrastive loss for generated-real IHC pairs. By replacing $L^{he}$ and $L^{gt}$ with $L^{he}_{mix}$ and $L^{gt}_{mix}$, the proposed method brings significant performance gains on FID, PHV, and KID, demonstrating the effectiveness of MDCL.

\begin{table}
\caption{The ablation study on HER2 by replacing the original patch-based contrastive loss with MDCL. KID values are shown by multiplying 1000.}\label{tab_abl}
\centering
\begin{tabular}{|l|c|c|c|}
\hline
 \multirow{2}*{Method} & \multirow{2}*{FID$\downarrow$} & \multicolumn{1}{c|}{PHV$_{T=0.01}$$\downarrow$} & \multirow{2}*{KID$\downarrow$}\\
& &  avg.(layer1$\sim$4)& \\
\hline
$L^{he}$ & 62.5 & 0.4997&  26.9\\
$L_{mix}^{he}$ & {\bfseries 59.4} &  {\bfseries 0.4960} &{\bfseries 21.1}\\
\hline
$L^{he}+L^{gt}$ & 51.4 & 0.4881 & 12.4\\
$L_{mix}^{he}+L_{mix}^{gt}$ & {\bfseries 44.4} &  {\bfseries 0.4751} &{\bfseries7.5}\\
\hline
\end{tabular}
\end{table}

\begin{figure*}[!t]
\centering
\includegraphics[width=0.85\textwidth]{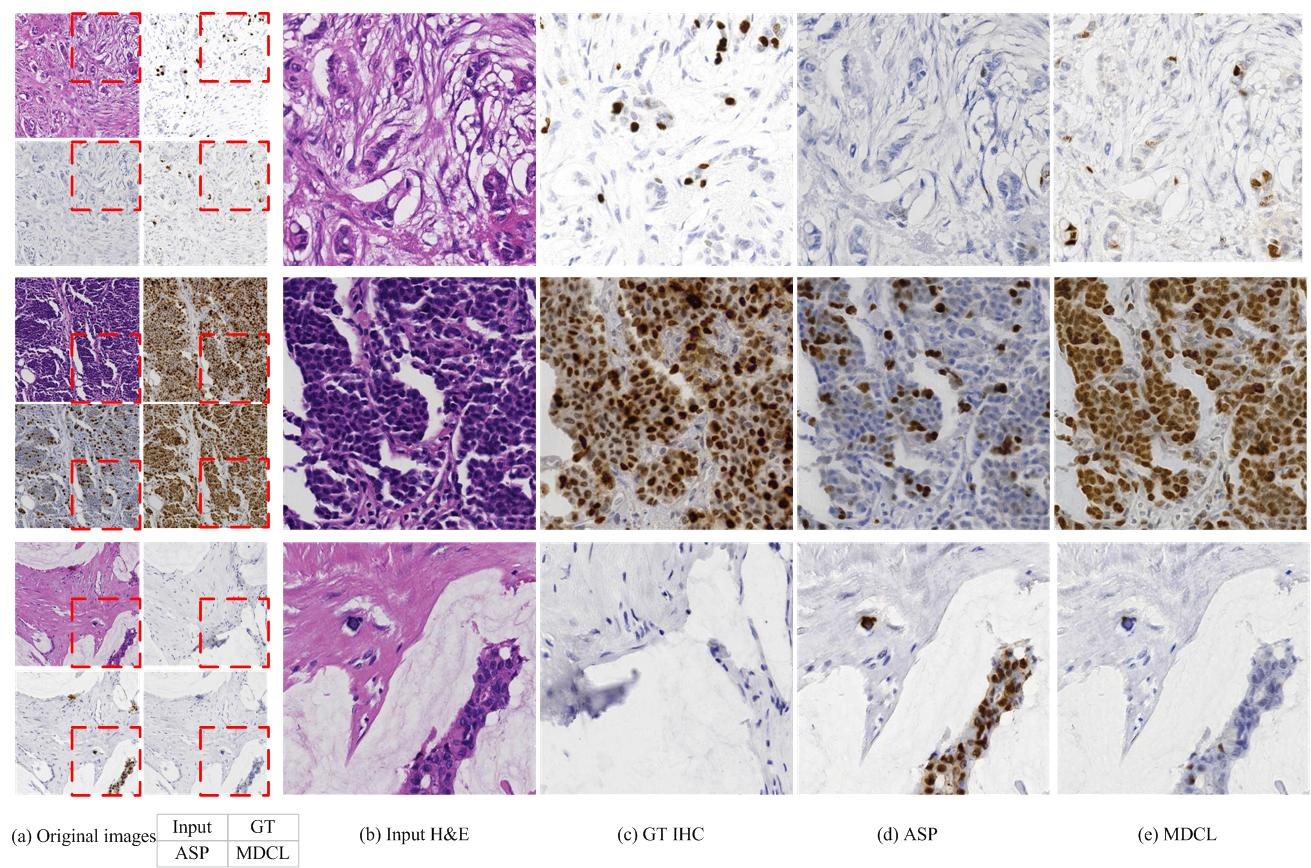}\vspace{-3pt}
\caption{Visualization results of some H\&E-IHC image pairs. (a) Original images, including input H\&E images, GroundTruth IHC images, and the generated IHC images with ASP and MDCL. (b) Input H\&E patch, (c) Groundtruth IHC patch, (c) Generated IHC patch with ASP, (d) Generated IHC patch with MDCL.} \label{visual}
\end{figure*}

\noindent\textbf{Visualization Results}. Some visual comparison results are illustrated in Fig.\ref{visual}. Fig.\ref{visual} (a) shows the images with the original size. To highlight the visual effect of the generated IHC images, we crop the local patches in the red boxes and list them as (b) Input H\&E, (c) GT IHC, (d) ASP, and (e) MDCL. Compared with the baseline ASP, MDCL generates the translation results with more discriminative component details. For example, in the first two rows of Fig.\ref{visual}, the positive tumor cells (brown) are accurately translated by MDCL whereas the virtual IHC images generated by ASP contain rare positive tumor cells. When translating the negative tumor cells into the IHC domain (the last row), ASP generates incorrect staining results. Meanwhile, the generated IHC image with MDCL has an extreme similarity with the groundtruth.

\subsection{Experiment on BCI}
The superiority of the proposed method is also validated on BCI dataset. Since the training and testing sets in \cite{li2023adaptive} are different from the default ones, we retrained the ASP model to make a fair comparison with our method. The evaluation results on BCI are illustrated in Table \ref{tab_bci}. It is evident that our model achieves the best performance gains on all metrics, demonstrating the robustness of the proposed MDCL method.

\begin{table}[!t]
\small
\caption{Evaluation results on BCI. 
KID values are shown by multiplying 1000. The symbol $^*$ means that we retrain the model on the default training set from BCI.}\label{tab_bci}
\centering
\begin{tabular}{|l|c|cccc|c|}
\hline
 \multirow{2}*{Method} & \multirow{2}*{FID$\downarrow$} & \multicolumn{4}{c|}{PHV$_{T=0.01}$$\downarrow$} & \multirow{2}*{KID$\downarrow$}\\
& & layer1 & layer2 & layer3 & layer4 & \\
\hline

ASP$^*$ & 219.4 & 0.6436& 0.5761& 0.3860& 0.8297 &   95.5\\
MDCL & {\bfseries 51.2} & {\bfseries 0.4962} & {\bfseries 0.3861} & {\bfseries 0.2351} & {\bfseries 0.7344} &{\bfseries 13.0}\\
\hline
\end{tabular}
\end{table}

\section{Conclusion}
In this paper, we have proposed the mix-domain contrastive learning for unpaired H\&E-to-IHC stain translation. The proposed method treats matching images as different domains, and formulates a mix-domain contrastive loss, accounting for the impact of negative patches from both inter- and intra-domain on pathological consistency learning. Experimental results demonstrate that the proposed method achieves state-of-the-art performance on two publicly available H\&E-to-IHC translation datasets.

\bibliographystyle{IEEEbib}
\bibliography{mybib}

\end{document}